\documentclass[journal]{IEEEtran}
\usepackage{times}
\usepackage{soul}
\usepackage{url}
\usepackage[hidelinks]{hyperref}
\usepackage[utf8]{inputenc}
\usepackage[small]{caption}
\usepackage{amsmath}
\usepackage{amsthm}
\usepackage{booktabs}
\usepackage{algorithm}
\usepackage{algorithmic}
\usepackage{enumitem}
\usepackage{subfigure}
\usepackage{multirow}
\usepackage{makecell}

\usepackage{scalerel}
\usepackage{tikz}
\usetikzlibrary{svg.path}

\definecolor{orcidlogocol}{HTML}{A6CE39}
\tikzset{
  orcidlogo/.pic={
    \fill[orcidlogocol] svg{M256,128c0,70.7-57.3,128-128,128C57.3,256,0,198.7,0,128C0,57.3,57.3,0,128,0C198.7,0,256,57.3,256,128z};
    \fill[white] svg{M86.3,186.2H70.9V79.1h15.4v48.4V186.2z}
                 svg{M108.9,79.1h41.6c39.6,0,57,28.3,57,53.6c0,27.5-21.5,53.6-56.8,53.6h-41.8V79.1z M124.3,172.4h24.5c34.9,0,42.9-26.5,42.9-39.7c0-21.5-13.7-39.7-43.7-39.7h-23.7V172.4z}
                 svg{M88.7,56.8c0,5.5-4.5,10.1-10.1,10.1c-5.6,0-10.1-4.6-10.1-10.1c0-5.6,4.5-10.1,10.1-10.1C84.2,46.7,88.7,51.3,88.7,56.8z};
  }
}

\newcommand\orcidicon[1]{\href{https://orcid.org/#1}{\mbox{\scalerel*{
\begin{tikzpicture}[yscale=-1,transform shape]
\pic{orcidlogo};
\end{tikzpicture}
}{|}}}}

\newtheorem{mydef}{Definition}
\newtheorem{myprob}{Problem}

\newcommand{\Graph}{\mathcal{G}}
\newcommand{\POISET}{\mathcal{P}}
\newcommand{\USERSET}{\mathcal{U}}
\newcommand{\EDGESET}{\mathcal{E}}
\newcommand{\SEQUENCE}{\mathcal{S}}
\newcommand{\Function}{\mathcal{F}}
\newcommand{\History}{\mathcal{H}}
\newcommand{\Loss}{\mathcal{L}}

\newcommand{\xvec}{\mathbf{x}}
\newcommand{\hvec}{\mathbf{h}}
\newcommand{\evec}{\mathbf{e}}
\newcommand{\Emat}{\mathbf{E}}
\newcommand{\svec}{\mathbf{s}}
\newcommand{\Smat}{\mathbf{S}}
\newcommand{\vvec}{\mathbf{v}}
\newcommand{\Vmat}{\mathbf{V}}
\newcommand{\Wmat}{\mathbf{W}}

\newcommand{\LSTM}{\mathrm{LSTM}}
\newcommand{\Precision}{\mathrm{P@}}
\newcommand{\PrecisionK}{\mathrm{P@k}}

\newcommand{\Recall}{\mathrm{R@}}
\newcommand{\RecallK}{\mathrm{R@k}}

\ifCLASSINFOpdf
\else
\fi
\begin{document}

\title{Social Link Inference via Multi-View Matching Network from Spatio-Temporal Trajectories}
%\href{https://orcid.org/0000-0001-6763-8146}
%\orcidicon{0000-0001-6763-8146}

\author{Wei~Zhang \orcidicon{0000-0001-6763-8146},~\IEEEmembership{Member,~IEEE,}
        Xin~Lai, and~Jianyong~Wang,~\IEEEmembership{Fellow,~IEEE}% <-this % stops a space
\thanks{Manuscript received xx; revised xxx; accepted xxx. Date of publication xxx; date of current version xxx.
This work was supported in part by the National Key Research and Development Program of China (No. 2019YFB2102600), in part by Shanghai Sailing Program (No. 17YF1404500), in part by the National Natural Science Foundation of China (No. 61702190, No. U1609220, No. 61532010, and 61521002), in part by the foundation of Key Laboratory of Artificial Intelligence, Ministry of Education, P.R. China, and in part by Beijing Academy of Artificial Intelligence (BAAI). 
\textit{(Corresponding author: Wei Zhang.)}
}
\thanks{W. Zhang is with the School of Computer Science and Technology, East China Normal University, Shanghai 200062, China, and also with the MOE Key Lab of Artificial Intelligence, Shanghai Jiao Tong University, Shanghai 200240, China (e-mail: zhangwei.thu2011@gmail.com).}

\thanks{X. Lai are with the School of Computer Science and Technology, East China Normal University, Shanghai 200062, China (e-mail: 51184501121@stu.ecnu.edu.cn).}% <-this % stops a space
\thanks{J. Wang is with the Department of Computer Science and Technology, Tsinghua University, Beijing 100086, China (e-mail: jianyong@tsinghua.edu.cn).}% <-this % stops a space
}


\markboth{Journal of IEEE TRANSACTIONS ON NEURAL NETWORKS AND LEARNING SYSTEMS,~Vol.~xx, No.~xx, xxxx}%
{Shell \MakeLowercase{\textit{et al.}}: Bare Demo of IEEEtran.cls for IEEE Journals}



\maketitle

\begin{abstract}
In this paper, we investigate the problem of social link inference in a target Location-aware Social Network (LSN), which aims at predicting the unobserved links between users within the network.
This problem is critical for downstream applications including network completion and friend recommendation.
In addition to the network structures commonly used in general link prediction, the studies tailored for social link inference in an LSN leverage user trajectories from the spatial aspect.
However, the temporal factor lying in user trajectories is largely overlooked by most of the prior studies, limiting the capabilities of capturing the temporal relevance between users.
Moreover, effective user matching by fusing different views, i.e., social, spatial, and temporal factors, remains unresolved, which hinders the potential improvement of link inference.
To this end, this paper devises a novel multi-view matching network (MVMN) by regarding each of the three factors as one view of any target user pair. 
MVMN enjoys the flexibility and completeness of modeling each factor by developing its suitable matching module: \romannumeral1) location matching module, \romannumeral2) time-series matching module, and \romannumeral3) relation matching module.
Each module learns a view-specific representation for matching, and MVMN fuses them for final link inference.
Extensive experiments on two real-world datasets demonstrate the superiority of our approach against several competitive baselines for link prediction and sequence matching, validating the contribution of its key components.

\end{abstract}


% \fi
\IEEEpeerreviewmaketitle

\section{Introduction}\label{sec:intro}
\IEEEPARstart{A}{s} the core part of fast-developing social medias, user linked data has been popularized unprecedentedly around the online world.
This makes link prediction~\cite{Liben-Nowell-CIKM2003}, aiming to predict the unobserved links between nodes (e.g., users) given some observed links in target networks, attractive to both academic and industrial domains.
Link prediction benefits a variety of downstream applications, such as network completion~\cite{Kim-SDM2011}, friend recommendation~\cite{Roth-SIGKDD2010}, event recommendation~\cite{ZhangW15}, to name a few.
Most existing studies dive into this problem from the perspective of learning from social network structures, for example, heterogeneous networks~\cite{Sun-ASONAM2011} and coupled networks~\cite{Dong-SIGKDD15}.
In particular, recent advances in representation learning on graphs~\cite{Hamilton-Arxiv2017,Cui-TKDE2018} significantly promote its development.

In location-aware social networks (LSNs), despite the common social links, users have rich spatio-temporal trajectories recorded by widely prevalent GPS-enabled mobile devices.
A typical spatial trajectory involves multiple sequential check-ins from a user, where each check-in is composed of a location and a timestamp.
The temporal and spatial properties of check-ins make the mobile trajectories substantially different from other types of sequential data. 
Since the spatio-temporal trajectories contain informative user mobility patterns, many studies learn to discover knowledge from them for different real applications, including location prediction~\cite{Liu-AAAI2016} and recommendation~\cite{Chang-IJCAI2018}, check-in time prediction~\cite{PanRAG16}, and trajectory-user linking~\cite{Gao-IJCAI2017}.
To sum up, all these studies are attributed to performing prediction for a single user each time.

%\begin{figure}[!t]
%\centering
%\begin{minipage}[c]{0.48\textwidth}
%\subfigure{\includegraphics[width=1\textwidth]{model-gen.eps}}
%\caption{The sketch of MVMN for social link inference.
%Three aspects of LSNs, i.e., spatial, temporal, and social factors, are %reflected by three subfigures connected with a user.}\label{fig:sketch}
%\end{minipage}
%\end{figure}

\begin{figure}[!t]
\centering
%\begin{minipage}[c]{0.48\textwidth}
%\subfigure{
\includegraphics[width=0.49\textwidth]{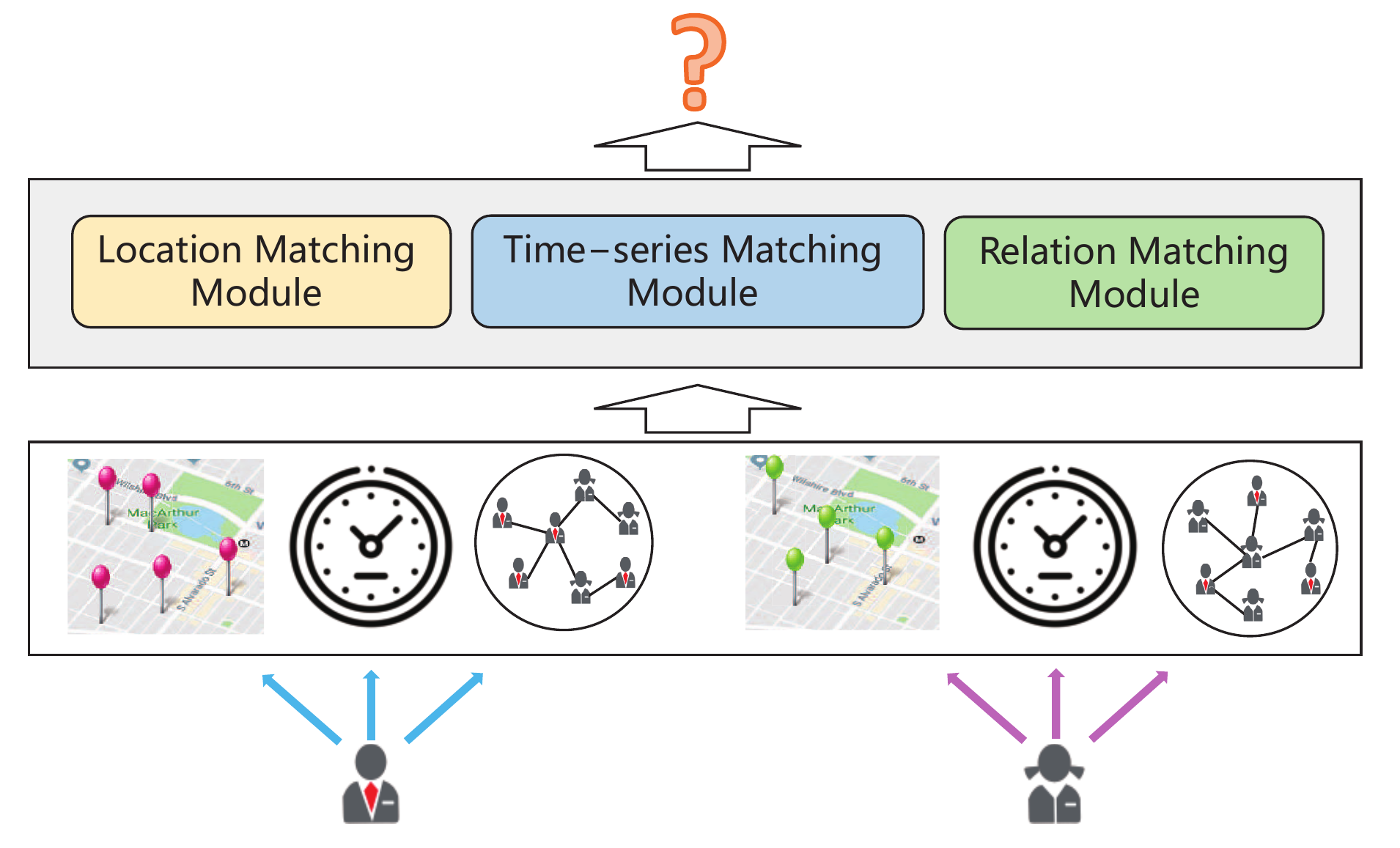}
\caption{The sketch of MVMN for social link inference.
Three aspects of LSNs, i.e., spatial, temporal, and social factors, are reflected by three subfigures connected with a user.}\label{fig:sketch}
%\end{minipage}
\end{figure}

In this paper, we investigate the problem of social link inference in a target LSN~\cite{Scellato-SIGKDD2011} which estimates whether a given user pair has a social link based on their trajectories.
This makes it fundamentally different from the above studies towards LSNs.
In comparison with general link prediction, user mobility from spatial trajectories is utilized in this problem.
Considerable efforts~\cite{Scellato-SIGKDD2011,HsiehL14,ZhangKY14} on this problem adopt feature based classification methods.
They require manually crafted mobility features from user trajectory pairs to train user relation classifiers.
The recent approaches~\cite{YangQYC19,WuLJC19} benefit from the powerful graph representation learning to encode both users and locations into low-dimensional vectors. They could be further utilized for deriving social relations.
However, most of them largely overlook the temporal aspect of user mobility.
This might cause information loss.  
It is intuitive that if two users visit the same physical location around the same time, there is a larger chance that they might have some interactions and establish a connection (see Table~\ref{tab:time_ana}).
Moreover, as shown later in Fig.~\ref{fig:temporal-similarities}, the users with social links have large temporal similarities than unlinked users.
As a result, the ignorance of temporal information would limit the accuracy of correlation modeling of any two user trajectories.
Although the study~\cite{YangQYC19} takes a step in this direction by discretizing continuous timestamps into time intervals and further treat them as nodes in graphs, the temporal information is not directly leveraged for linking users.
Moreover, the continuity and sequentiality of time are not well preserved by the intervals. 

It is worth noting that the problem of user identity linkage across different LSNs~\cite{Riederer-WWW2016,Chen-CIKM2017,Chen-ICDE2018,FengZWYZ0J19}, focusing on linking online IDs (existing in different social networks) of the same user, shares some similar spirits with the studied problem because of the consideration of user mobility traces.
Nevertheless, an essential difference is that there are more than one social network involved in user identity linkage.
This makes the manners of using network structures for the two problems fundamentally different.
For example, network alignment~\cite{Singh-PNAS2008,ZhangTYPY15,DuYZ19} techniques proposed for node linkage across networks are not applicable for the scenarios with only one target social network. 
As such, although the above methods could be partially adapted to our problem, the ways of utilizing social relations should be carefully reconsidered. 

In summary, all the previous methods lack the mechanisms of performing user linking in a target LSN by effectively fusing the multiple views --- social, spatial, and temporal views.
To this end, we devise a Multi-View Matching Network (MVMN) and make the main contributions as follows:

\romannumeral1) We address the issue of overlooking the temporal factor in social link inference through detailed data analyses of the temporal influence on this problem.
Motivated by this, we develop a time-series module composed by a loosely coupled neural temporal point process.
It is benefited from the powerful temporal point process~\cite{Durbin1971Spectra}, with a simple extension of loosely coupling two Recurrent Neural Networks (RNNs) in Recurrent Marked Temporal Point Process (RMTPP)~\cite{DuDTUGS16}.
It welcomed for the capability of learning to match any pair of user time-series by capturing their continuity and sequentiality.
To our best knowledge, this is the first work to adopt point process for learning to match user trajectories, which might benefit the application studies of temporal point process in the literature.

\romannumeral2) Our proposed model MVMN learns the user-to-user correlation based on its multiple views, which is welcomed for the adequate utilization of different information types.
The sketch of MVMN is shown in Fig.~\ref{fig:sketch}, wherein each view corresponds to a tailored module in MVMN to contribute to correlation computation.
Despite the above time-series module, a location matching module uses a pairwise matching matrix to measure the relevance of each location pair.
Meanwhile, a relation matching module employs graph neural networks to characterize the proximity between users.
The learned multiple view-specific representations for matching are seamlessly fused for final link inference.
Such a design provides flexibility and completeness of modeling each view more effectively. 

\romannumeral3) Experimental results on real public datasets demonstrate the superiority of our approach over several strong competitors, including general link prediction methods (e.g., GAT~\cite{Velickovic-1710-10903}), link inference approaches for LSNs (e.g., Heter-GCN~\cite{WuLJC19}), user identity linkage methods (e.g., DPLink~\cite{FengZWYZ0J19}), and sequence matching models (e.g., ESIM~\cite{ChenZLWJI17}).
Ablation studies show the effectiveness of each module in MVMN.
In particular, the incorporation of RMTPP into our model, accompanied by temporal loss functions, has been verified to indeed contribute to inference performance.
The model source code\footnote{\url{https://github.com/runnerxin/MVMN}} is released for relevant research.

\section{Related Work}\label{sec:related}
In this section, we review the literature from three aspects: social link inference, spatio-temporal trajectory mining, and the core background techniques.

\subsection{Social Link Inference}
The history of link inference in social networks is not very long, dating back to the early study~\cite{Liben-Nowell-CIKM2003} which investigates how the different network based measures of node proximity affect inference performance.
Since then, link prediction has been extensively studied.
On the methodological aspect, various approaches like supervised random walk~\cite{Backstrom-WSDM2011}, factor graph~\cite{Hopcroft-CIKM2011}, and deep learning~\cite{Li-ICDM2014}, have been applied to the problem.
On the other hand, different categories of social networks have been explored, covering heterogeneous networks~\cite{Sun-ASONAM2011} having more than one type of nodes and edges, signed social networks~\cite{Leskovec-WWW2010} having positive and negative edges, and coupled networks~\cite{Dong-SIGKDD15} whereby more than one social network are considered.

In contrast to the above studies focusing on network structures, some other studies additionally leverage spatio-temporal trajectories of users to perform link inference.
The seminal work~\cite{Scellato-SIGKDD2011}, as a representative of feature based classification methods~\cite{Wang-SIGKDD2011,HsiehL14,ZhangKY14}, designs hand-crafted features such as the co-occurrence of visited locations and overlapping ratios of social friends.
The study~\cite{LuQZHG16} jointly factorizes the user-location interaction matrix, user-event interaction matrix, and social relation interaction matrix through the Bayesian latent factor model for friend recommendation.
The recent studies leverage the power of graph representation learning to obtain low-dimensional representations of users and locations~\cite{WuLJC19}, as well as time intervals~\cite{YangQYC19}.
However, only user representations are directly leveraged for user matching, while the spatial and temporal views of user matching are not fully exploited. 
In a nutshell, most of the above studies have overlooked the temporal aspect of user trajectories.
Although the study~\cite{YangQYC19} regards timestamps as nodes to learn the representations, the discretization of timestamps misses the continuity and sequentiality of time.
In this paper, we regard spatial, temporal, and social factors as different views of user pairs and perform view-specific user matching to sufficiently utilize the three types of information.

\subsection{User Spatio-temporal Trajectory Mining}\label{sec:trajectory}
User spatio-temporal trajectory mining is a hot research theme for understanding user mobility patterns.
Despite the aforementioned research on user link prediction with spatial trajectories, a variety of other applications have been investigated~\cite{Zheng-TIST2015}, wherein recent advances have abundantly utilized sequential representation learning.
Liu et al.~\cite{Liu-AAAI2016} developed context-aware RNNs for the next location prediction.
Chang et al.~\cite{Chang-IJCAI2018} conducted successive location recommendation by combining location and content representation learning.
Moreover, the check-in time is estimated by incorporating the check-in representations learned by RNNs into survival analysis~\cite{Yang-IJCAI2017} and point process~\cite{LiangZW19}, respectively.
Similarly, RNNs are employed to learn trajectory representation~\cite{Gao-IJCAI2017}, used as the input to a softmax function for calculating the probabilities of all candidate users to be linked.
In computer vision, some studies propose spatial-temporal image networks for video super-resolution~\cite{ZhuLZLLX19}, action recognition and localization~\cite{ZhuXAAAI20}.
However, as introduced in Section~\ref{sec:intro}, all the above problems take only one (user) sequence as input each time to estimate its target variable.
This makes their problem settings substantially different from the studied user link inference problem.

User identity linkage across different LSNs is a more similar problem setting.
It links accounts of the same user from different social networks based on their spatio-temporal trajectories
Specifically, the spatial and temporal spaces are divided into different time intervals\cite{Riederer-WWW2016}.
The co-occurrence of two users in the same interval determines their similarities.
Chen et al. first considered to measure the similarity of stay regions, local and global time distributions~\cite{Chen-CIKM2017}, and then further improved accuracy using kernel density estimation~\cite{Chen-ICDE2018}.
The study~\cite{FengZWYZ0J19} is the first one to use deep learning approaches in this problem.
RNNs are leveraged to encode a pair of user trajectories and an attention based selector is proposed to capture the correlations between the two trajectories.
However, this problem involves multiple social networks and only concentrates on linking the same user across networks.
Hence the consideration of social relations is fundamentally different and the existing user relations are commonly not used in these studies.

\subsection{Core Background Techniques}
\textbf{Temporal point process}~\cite{Durbin1971Spectra} is a principled approach for handling temporal sequences in a continuous space.
It has been applied to paper citation count modeling~\cite{XiaoYLJWYCZ16}, tweet popularity modeling~\cite{zhao2015seismic}, merger and acquisition activity prediction~\cite{YanXLJWKYZ16}, and patient flow prediction~\cite{xu2016patient}, to name a few.
More recently, point process also finds its application in spatio-temporal domain~\cite{OkawaIK0TU19}.
However, few relevant studies have considered applying it to sequence matching tasks.
To capture the continuity and sequentiality of time, we simply extend RMTPP~\cite{DuDTUGS16}, a powerful recurrent neural point process model, to a loosely coupled version to enable the simultaneous modeling of two time-series.

\textbf{Graph neural networks} have emerged in recent years as effective tools for encoding high-order relations of graphs into low-dimensional node embeddings~\cite{ScarselliGTHM09,Hamilton-Arxiv2017}.
In particular, graph convolution network (GCN)~\cite{DefferrardBV16} and graph attention network (GAT)~\cite{Velickovic-1710-10903} are two representatives.
GraphSAGE~\cite{HamiltonYL17} leverages node features such as text attributes for achieving inductive graph representation learning.
For the reason that GAT can automatically determine the weights of user neighbors, we adopt it to model social relations.

\textbf{Multi-view learning}
~\cite{xu2013survey,ZhaoXXS17}
is a paradigm exploiting the different views of the same input data to acquire abundant information for learning.
It has a variety of applications, e.g., learning view-specific embedding for clustering~\cite{YinWW18} and leveraging multi-view observations as supervisory signals for pose prediction~\cite{TulsianiEM18}.
In particular, the studies~\cite{ElkahkySH15,PalomaresBD18,WuLJC19} investigate the effectiveness of multi-view learning for recommendation, wherein the studies~\cite{ElkahkySH15,WuLJC19} regard each item domain as one view or take different parts of news as their different views. 
By contrast, each view in our study corresponds to one type of matching between two users.
As far as we know, this paper is the first study of leveraging multi-view learning for social link inference.

\section{Preliminaries}\label{sec:problem}

\subsection{Problem Formulation}
Let $\Graph=(\USERSET, \EDGESET, \POISET)$ represent a target location-aware social network to be investigated.
It involves three fundamental elements: (1) $\USERSET$ is a set of user IDs in the network, (2) $\EDGESET$ denotes the known social links between users, and (3) $\POISET$ represents a set of locations IDs corresponding to POIs in the real world.
In this type of social network, one of the main user behaviors is check-ins defined as follows:
\begin{mydef}[Check-in]
	A check-in $c$ is a tuple w.r.t. a location $l\in\POISET$ and a timestamp $t$, i.e., $c=(l, t)$.
\end{mydef}

Given any user ID $u\in\USERSET$, we define its check-in trajectory as follows:

\begin{mydef}[Check-in Trajectory]
	The check-in trajectory of user $u$ is constituted by multiple consecutive check-ins denoted as $\SEQUENCE_u=\{c^u_{1},\ldots,c^u_{\ell(u)}\}$, where $\ell(u)$ represents the number of check-ins.
\end{mydef}

Given user check-in trajectories and existing social links, the aim of social link inference in an LSN is to estimate the existence of social links between users which currently do not belong to $\EDGESET$.
We formulate the studied problem as follows:
\begin{myprob}[Social Link Inference in an LSN]
	Given a pair of users $u_m$ and $u_n$, the goal of this problem is to learn a model
	$\Function(\SEQUENCE_{u_m}, \SEQUENCE_{u_n}, \EDGESET)\to \hat{y}\in\{0, 1\}$ based on existing user trajectories and social links, which can accurately estimate whether they have a social link $(1)$ or not $(0)$.
\end{myprob}
The above problem setting illustrates that the key to success is effective model design and training.

\subsection{Motivation of Temporal Influence on Link Inference}\label{sec:motivation}

\begin{figure}[!t]
\centering
\includegraphics[width=.47\textwidth]{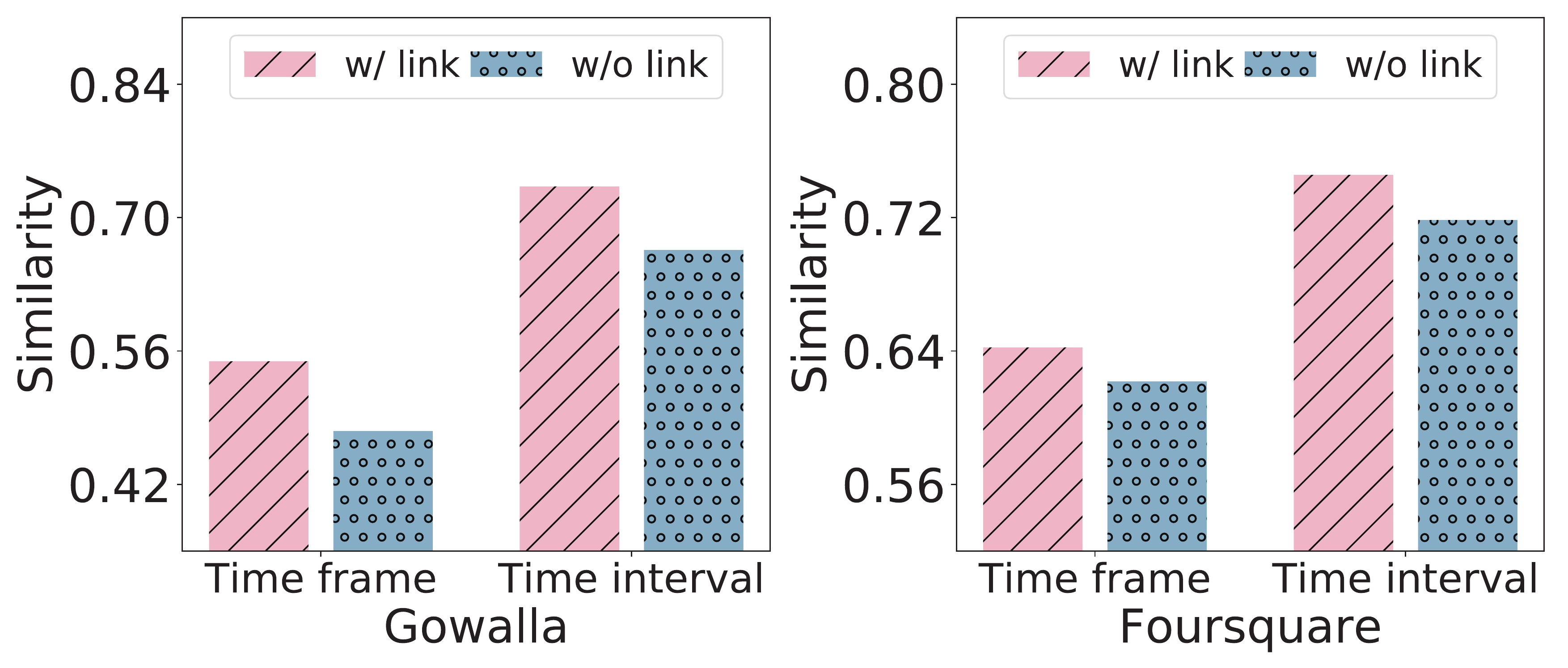}
\caption{Comparison of temporal similarities.}\label{fig:temporal-similarities}
\end{figure}

In this part, we carry out preliminary data analyses on Gowalla and Foursquare (shown in Section~\ref{sec:dataset}) to validate the rationality of considering the temporal influence on social link inference.
The analyses involve two aspects of comparisons, i.e., temporal similarities of linked users versus temporal similarities of unlinked users and ratios of spatial co-occurrence versus ratios of spatio-temporal co-occurrence for linked users:

\begin{enumerate}[leftmargin=*]
    \item  We define two types of temporal similarities to realize the first comparison, i.e., time frame based similarities and time interval based similarities.
    Time frames indicate the time when user check-in behaviors happen, while time intervals show the temporal difference between two consecutive check-in behaviors.
    Specifically, we have 24 time frames, each of which corresponds to one hour per day, to denote the frame based check-in distributions of users' behaviors.
    Consequently, the time frame based similarities are calculated by the average cosine similarities of linked users (w/ link) or unlinked users (w/o link).
    Without loss of generality, we further define 7 time intervals in hours: $[0,1), [1,2), [2,6), [6,12), [12, 24)$, and $[24, \infty)$, which are utilized to represent interval based check-in distributions.
    Similar to the time frame based similarities, we also use average cosine similarities to obtain time interval based similarities.
    Fig.~\ref{fig:temporal-similarities} compares the similarities for linked users (w/ link) or unlinked users (w/o link).
    It is clear that linked users have larger temporal similarities than unlinked users.

    \item Ratios of spatial co-occurrence versus ratios of spatio-temporal co-occurrence for linked users are calculated by the following manner. 
    Assume the number of user pairs visiting at least one common location is $N^L$, wherein the count of user pairs also having social links is $N^{L}_S$.
    Spatial ratio computation is given as: $SR = \frac{N^{L}_S}{N^L}$.
    We further compute spatio-temporal ratio in a similar fashion, by denoting the number of user pairs visiting at least one common location in the same time interval (the length of time interval is set to 1 hour) is $N^{LT}$, and the corresponding number of user pairs having social links as well is $N^{LT}_S$.
    Based on the two counts, spatio-temporal ratio is calculated by: $STR = \frac{N^{LT}_S}{N^{LT}}$.
    Table~\ref{tab:time_ana} compares the two types of ratios, from which we observe the additional consideration of temporal information could bring benefits to the social relation refinement.

\end{enumerate}

In summary, the above analyses demonstrate that the temporal aspect of user check-in behaviors has a great effect on social relations, which motivates us to utilize the powerful temporal point process to capture this for boosting the performance of social linkage inference.

\section{The Computational Methodologies}\label{sec:model}
\begin{figure*}[!htbp]
    \centering
        \includegraphics[width=.97\textwidth]{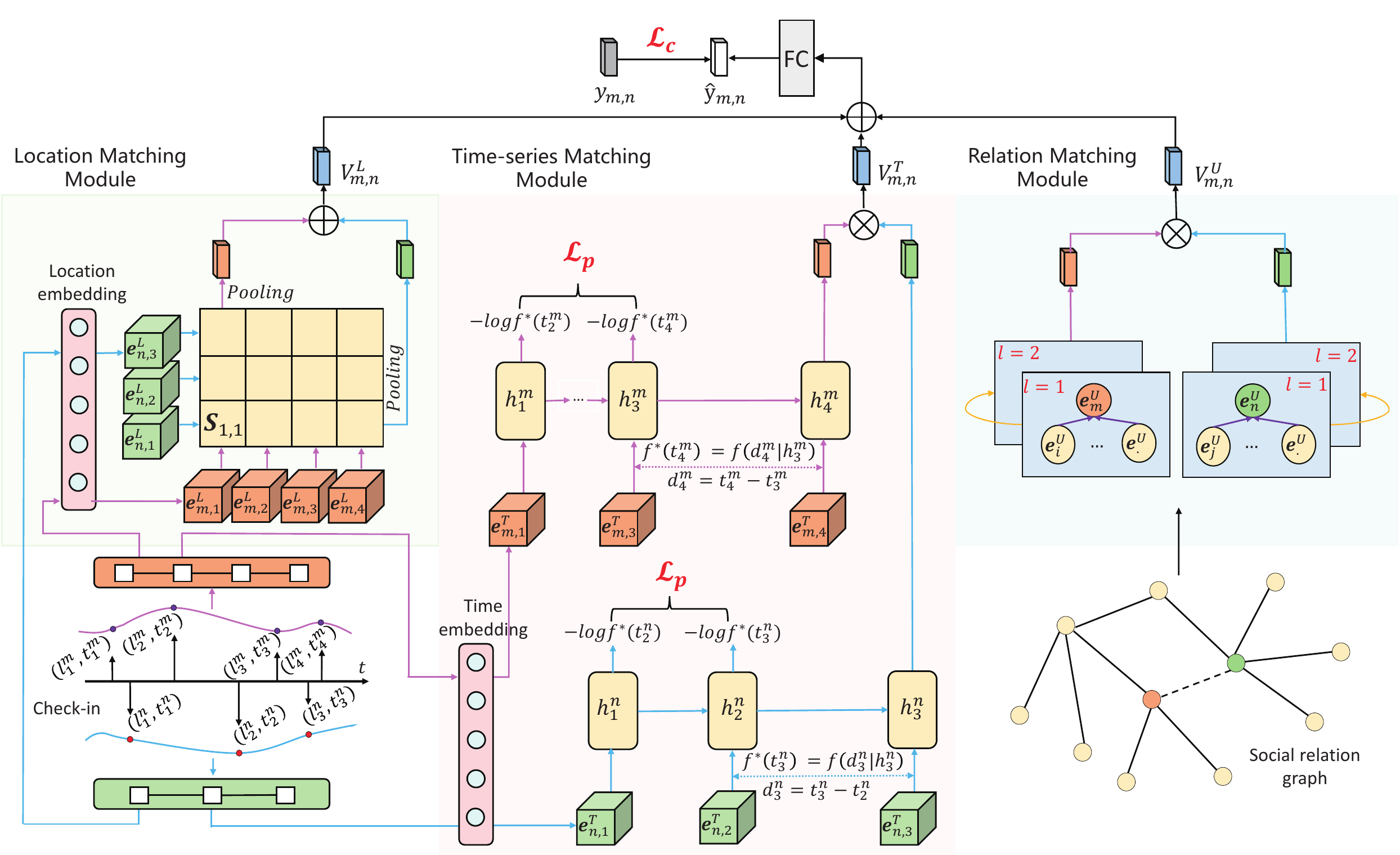}
        \caption{The architecture of MVMN. We visualize it with two toy trajectory examples. One is for user $u_m$ with length 4 and the other is for user $u_n$ with length 3. 
        The three color regions correspond to different modules. The blue and pink arrows denote the information flow of different users. The loss functions are marked with red font.}\label{fig:model}
\end{figure*}

The architecture of our model MVMN is presented in Fig.~\ref{fig:model}.
The input for MVMN is composed of a user pair with check-in trajectories and a social relation graph containing the existing user links.
The output of the model is the estimation of the possibility that the user pair has a social connection.
In MVMN, there are three major modules: \textit{location matching module}, \textit{time-series matching module}, and \textit{relation matching module}.
The three modules provide different views to characterize the degree of matching between two users. 
In the end, the matching based representations from multiple views are integrated to calculate the relevance score of the two given users.
The decomposition of relevance computation into different views enables the flexibility of designing tailored methods for each view.

\begin{table}[!t]
	\centering
	\caption{Ratio comparison for spatial co-occurrence and spatio-temporal co-occurrence. }\label{tab:time_ana} 	
	\begin{tabular} {c|cc} 
		\toprule   
		Dataset & spatial ratio  & spatio-temporal ratio \\		 
        \midrule  
         Gowalla     & 35.8\%  & 73.4\% \\
	   Foursquare    & 26.8\%  & 52.6\% \\
		\bottomrule   
	\end{tabular}  
\end{table}

\subsection{Input Representation}
We encode locations, timestamps, and users into low-dimensional vector spaces for ease of later representation learning. 
Specifically, the one-hot encoding $\xvec_u$ of user $u$ can be first acquired by its ID. 
Afterwards, a commonly adopted lookup operation is performed, i.e., $\evec^U_u = \Emat_U^{\top}\xvec_u$.
Similarly, location $l$ is converted into $\evec^L_l$ based on its ID and embedding matrix $\Emat_L$ as well. 
It is a little different for dealing with timestamps due to its continuity.
Here we first divide each timestamp into different time intervals, and later will use temporal point process for modeling continuous timestamps.
We also encode each time interval by the usage of the time embedding matrix.
Thus timestamp $t$ is represented as $\evec^T_t$.

Without loss of generality, we take $u_m$ and $u_n$ as an example for specifying the matching details of MVMN.
The spatial, temporal, and social aspects are distinguished by different views of the user pair.
In particular, user $u_m$ is associated with a spatial sequence $L_m=\{\evec^{L}_{m,1}, \ldots, \evec^{L}_{m,\ell(u_m)}\}$, a temporal sequence $T_m = \{\evec^{T}_{m,1}, \ldots, \evec^{T}_{m,\ell(u_m)}\}$, and a social representation $\evec^U_{m}$.
Likewise, it is the same for user $u_n$ to have its corresponding representations.

\subsection{Location Matching Module}
From the geographical view of the user trajectory pair, we first calculate a pairwise matching matrix to measure the correlation between one location in the spatial sequence of user $u_m$ and another location related to user $u_n$.
Formally, given the $i$-th location in $L_m$ and the $j$-th location in $L_n$, the cosine similarity is computed as follows:
\begin{equation}\label{eq:cosine}
\Smat_{i, j} = \cos(\evec^{L}_{m,i}. \evec^{L}_{n,j})\,.
\end{equation}
After traversing all pairs of locations in the two sequences, we can obtain the similarity matrix $\Smat$ where each entry corresponds to one location pair.

It is intuitive that if the two spatial sequences are more similar, there is a bigger chance that the two users will have a social connection.  
With this intuition, we perform row-wise max-pooling and column-wise max-pooling computations as:
\begin{equation}\label{eq:row-wise}
\svec^m_{i}=\max _{a \in\{1, \ldots, \ell(u_n)\}} \Smat_{i,a}~~~~~~i\in\{1, \ldots, \ell(u_m)\}\,,
\end{equation}
\begin{equation}\label{eq:column-wise}
\svec^n_{j}=\max _{a \in\{1, \ldots, \ell(u_m)\}} \Smat_{a,j}~~~~~~j\in\{1, \ldots, \ell(u_n)\}\,.
\end{equation}
The above equations return the largest relevance score for each location, which plays a role in summarizing the interaction of the two trajectories.

The two obtained relevance vectors are combined together to form the representation $\Vmat^L_{m,n} = [\svec^m\oplus\svec^n]$ to capture the correlation of user spatial trajectories, where $\svec^m = \{\svec^m_{1},\cdots,\svec^m_{\ell(u_m)}\}$ and $\svec^n = \{\svec^n_{1},\cdots,\svec^n_{\ell(u_n)}\}$.
$\oplus$ represents the concatenation operation.
In practice, we set a maximal length ($K$) of a trajectory to ensure the embedding size is fixed, which facilitates the later adoption of fully-connected layers to yield the final relevance score.

\subsection{Time-series Matching Module}
The temporal aspect of user trajectories is reflected in the continuity and sequentiality of time. 
As mentioned before, we propose loosely coupled RMTPP to learn time-series matching.
Specifically, we first feed the pair of temporal sequences $T_m=\{\evec^{T}_{m,1}, \ldots, \evec^{T}_{m,\ell(u_m)}\}$ and $T_n=\{\evec^{T}_{n,1}, \ldots, \evec^{T}_{n,\ell(u_n)}\}$ into recurrent neural networks which are implemented by LSTM~\cite{Hochreiter-NC97} for its better performance in our experiments. 
We omit the computational details of LSTM and define the brief formulas for obtaining sequential hidden representations as follows:
\begin{align}\label{eq:lstm1}
\hvec^m_{i} &= \LSTM(\evec^{T}_{m, i}, \hvec^m_{ i-1})~~~~i\in\{1, \ldots, \ell(u_m)\}\,,\\
\hvec^n_{j} &= \LSTM(\evec^{T}_{n, j}, \hvec^n_{ j-1})~~~~j\in\{1, \ldots, \ell(u_n)\}\,,
\end{align}
where $\LSTM(\cdot)$ corresponds to each time step. 

We adopt the finally returned embeddings $\hvec^m_{ \ell(u_m)}$ and $\hvec^n_{\ell(u_n)}$ to represent each user's time-series.
Given this, the correlation representation of user time-series is computed as follows:
\begin{equation}\label{eq:time-match}
\Vmat^T_{m,n} = \tanh(\hvec^m_{\ell(u_m)}\otimes\hvec^n_{ \ell(u_n)})\,,
\end{equation}
where $\otimes$ is the Hardmard product.
The usage of the activation function tanh($\cdot$) constrains the value in each dimension of $\Vmat^T_{m,n}$ in the range of [-1, 1].
As such, they are consistent with the cosine similarities adopted in the location matching module, which are more interpretable.
We have also tried some advanced techniques such as attention mechanisms~\cite{Bahdanau-Arxiv14} to fuse multiple hidden representations.
In particular, the attention computation is conducted over all the hidden representations and the query representations are obtained by their mean-pooling.
Yet the results show no significant improvements. 

In this way, the sequentiality of user time-series could be captured.
To further model the continuity of time to guide the effective representation learning of $\hvec^m_{ \ell(u_m)}$ and $\hvec^n_{\ell(u_n)}$, we introduce a temporal point process based loss for time-series.
In temporal point process, the conditional intensity function $\lambda^*(t)$ plays a key role in capturing the temporal dynamics in a continuous space.
By convention, * in the function denotes it is history dependent. 
Within a short time interval $[t,t+dt)$, $\lambda^*(t)$ represents the occurrence rate of a new check-in event given the history $\History_t$ and satisfies:  $\lambda^*(t)\,dt=P\{c \in [t,t+dt)|\History_t\}$. 
Based on this, the density function is given as:
\begin{equation}\label{eq:cdf}
f^*(t)=\lambda^*(t)\exp\Big(-\int_{t_j}^{t}{\lambda^*(\epsilon)\,d\epsilon}\Big)\,,
\end{equation}
where $t_j$ could be a happened timestamp of the last check-in event or a starting timestamp. 

In RMTPP, $\lambda^*(t)$ has a well-designed parametric form associated with the accumulative influence of past check-ins, the temporal gap since the last event happened, and a global base intensity to denote the context-independent occurrence of a next event.
For the given two users, their conditional intensity functions are defined as follows:
\begin{align}\label{eq:con-intensity}
\lambda_m^{*}(t) &= \exp\Big(\vvec^\top \hvec^m_{ i}+\omega\left(t-t^m_{i}\right) + b\Big)\,, \\
\lambda_n^{*}(t) &= \exp\Big(\vvec^\top \hvec^n_{ j}+\omega\left(t-t^n_{j}\right) + b\Big)\,,
\end{align}
where $\vvec$, $\omega$, and $b$ are trainable parameters. 
Through Eq.~\ref{eq:time-match}, the above two functions of recurrent temporal point process are loosely coupled.

Due to this simple $\lambda^*(t)$, the density function has an analytical form.
Here we clarify the density function $f_m^{*}(t)$ for user $u_m$:
\begin{equation}\label{eq:density}
\begin{array}{l}
{f_m^{*}(t)=\lambda_m^{*}(t) \exp \left(-\int_{t^m_{i}}^{t} \lambda_m^{*}(\epsilon) \,d\epsilon\right)} \\ 
{=\exp \left\{\vvec^\top \hvec^m_{ i}+\omega\left(t-t^m_{i}\right) + b+\frac{1}{\omega} \exp \left(\vvec^\top \hvec^m_{i}+b\right)\right.} \\
{\left.-\frac{1}{\omega} \exp \left(\vvec^\top \hvec^m_{i}+\omega\left(t-t^m_{i}\right)+b\right)\right\}\,.}
\end{array}
\end{equation}
Similarly, $f_n^{*}(t)$ could be derived.

In the end, the loss function of loosely coupled RMTPP is denoted as the negative joint log-likelihood of generating the time-series:
\begin{equation}\label{eq:pp-loss}
\Loss_p = -\sum_{i=1}^{\ell(u_m)-1}\log f_m^{*}(t^m_{i+1}|\hvec^m_{i})-\sum_{j=1}^{\ell(u_n)-1}\log f_n^{*}(t^n_{j+1}|\hvec^n_{j})\,.
\end{equation}
Through the optimization towards $\Loss_p$, the hidden representations $\hvec^m_{ \ell(u_m)}$ and $\hvec^n_{\ell(u_n)}$ could receive useful signals distilled from the continuous timestamps of check-ins.

\subsection{Relation Matching Module}
As discussed previously, we adopt GAT to learn user embeddings in the view of their social relations. 
The core advantage of GAT is the ability to determine the importance weights of neighbor nodes for each target user node.
This is appropriate for our study since the original binary values of social edges could not differentiate the relation strength between different users.
To be specific, the attention coefficient of the given user pair is computed as follows:
\begin{equation}\label{eq:coeff}
    c_{m,n} = \mathrm{FC}(\left[\Wmat \evec^U_{m} \oplus \mathbf{W} \evec^U_{n}\right])\,, 
\end{equation}
where FC($\cdot$) denotes one or more fully-connected layers.
$\Wmat$ is a parameter matrix for linear transformation.

Given the obtained attention coefficients, the importance weights of neighbors are normalized to be probability distributions for ease of updating representations later.
Taking user $u_m$ for illustration, we have the following formula to get the corresponding distribution:
\begin{equation}\label{eq:att-prob}
\alpha_{m, n}=\frac{\exp \Big(\mathrm{LeakyReLU}\left(c_{m,n}\right)\Big)}{\sum_{i \in \mathcal{N}_{m}} \exp \Big(\mathrm{LeakyReLU}\left(c_{m, i}\right)\Big)}\,,
\end{equation}
where $\mathcal{N}_{m}$ includes the user itself and its neighbor users.
$\alpha_{m,n}$ reflects the contribution of user $u_n$ to updating the representation for user $u_m$.
The updated representation is calculated by an aggregation function:
\begin{equation}\label{eq:aggregate}
\evec^{U}_{m}=\sigma\left(\sum_{i \in \mathcal{N}_{m}} \alpha_{m,i} \Wmat \evec^{U}_{i}\right)\,,
\end{equation}
where $\sigma(\cdot)$ is a nonlinear activation function (e.g., $\mathrm{ELU}(\cdot)$).
Besides, multi-header attention is commonly adopted as an extension of single-header attention to learn multiple attention coefficients to obtain multiple user node representations.
We take the average of these representations as the updated representations. 
More details can be referred to the study~\cite{Velickovic-1710-10903}.

We summarize the above basic procedures (from Eq.~\ref{eq:coeff} to Eq.~\ref{eq:aggregate}) for one-time graph propagation as $\mathrm{GAT}(\cdot)$.
Then it could be generalized to a multi-layer computational formula, given as:
\begin{equation}\label{eq:multi-layer-gat}
    \evec^U_{m,(k+1)} = \mathrm{GAT}(\Wmat, \evec^U_{m',(k)})~~~ m'\in\mathcal{N}_{m}\,,
\end{equation}
which corresponds to the representation propagation from layer $k$ to layer $k+1$.
For starting propagation, $\evec^U_{m,(0)}$ is set to $\evec^U_{m}$.

After $K$ layers of propagation, we obtain the final social representations $\evec^U_{m,(K)}$ and $\evec^U_{n,(K)}$ for the two users.
The Hardmard operation is leveraged as well to acquire the representation for social matching, defined as follows:
\begin{equation}\label{eq:social-match}
    \Vmat^U_{m, n} = \tanh(\evec^U_{m,(K)} \otimes \evec^U_{n,(K)})\,.
\end{equation}
The activation function tanh($\cdot$) plays a role similar to Eq.~\ref{eq:time-match}.

\subsection{Multi-view Fusion and Link Estimation}\label{subsec:multi-view}
In order to measure user-to-user correlations from a holistic perspective, our model integrates the view-specific representations by $[\Vmat^L_{m, n}\oplus\Vmat^T_{m, n}\oplus\Vmat^U_{m, n}]$.
The final social relation prediction is conditioned on the above integrated representation, which is defined as:
\begin{equation}
\hat{y}_{m,n} = \varphi\Big(\mathrm{FC}([\Vmat^L_{m, n}\oplus\Vmat^T_{m, n}\oplus\Vmat^U_{m, n}]) \Big)\,,   
\end{equation}
where $\varphi(\cdot)$ is a sigmoid function which outputs the probability of the user pair to have a social link.

The cross-entropy loss is adopted to quantify the gap between the predicted relation and ground-truth relation:
\begin{equation}\label{eq:ce_loss}
    \Loss_c = -\Big(y_{m,n} \log \hat{y}_{m,n} + (1-y_{m,n}) \log (1-\hat{y}_{m,n})\Big)\,, 
\end{equation}
where $y_{m,n}$ corresponds to the ground-truth relation label.
The above loss function is dedicated to the user pair of $u_m$ and $u_n$, and could be easily generalized to all other user pairs in the training stage.
Since for each user, the total number of its linked users in training data is relatively small compared to that of unlinked users, we sample a small number of unlinked users to build pseudo negative pairs.

The final objective function is a combination of the cross-entropy loss and the point process loss, which is written as:
\begin{equation}
    \Loss = \Loss_c + \beta \Loss_p\,,
\end{equation}
where $\beta$ controls the relative effect of the point process loss and could be tuned on the validation dataset.

\section{Experimental Setup}\label{sec:es}
To evaluate the performance of our model, we first illustrate basic experimental setups, including the datasets we used, the adopted evaluation metrics, the carefully chosen strong baselines, and some implementation details.

\subsection{Datasets}\label{sec:dataset}

We conduct experiments on two real-world and publicly available location-aware social networks:

\textbf{Gowalla\footnote{https://snap.stanford.edu/data/loc-Gowalla.html}:} 
This is a widely used dataset for relevant studies, collected by the work~\cite{cho2011friendship}.
The raw dataset consists of 196,591 users and 950,327 relations.
The number of corresponding check-ins is 6,442,890 crawled during the period of Feb. 2009 to Oct. 2010.
Although the dataset contains the sampled users over the planet, the relations mainly exist between two users in the same city due to the low possibility of  check-in behaviors happened around the world.
Therefore it is not very meaningful to conduct relation inference for all these users.
For this reason, we choose one of the largest cities, New York City, to select users whose check-ins are usually located in the city.
In particular, we first specify the region based on its longitude and latitude.
Consequently, we filter out those users with small proportions of her/his total check-ins (less than 0.1) within the specified region.
The subset selection by different cities in location-aware social networks is commonly used by the prior studies~\cite{ZhangWF13,YangQYC19}. 
To ensure the reliability of comparison, we remove users with less than one friend and ten check-ins.

\textbf{Foursquare\footnote{https://sites.google.com/site/dbhongzhi/}:} 
This dataset is crawled from the most popular LSN, i.e., Foursquare.
Similar to the data selection procedure for Gowalla, we choose another large city, Los Angeles City, to gather training data, which ensures data diversity. 
We follow the same preprocessing methods applied for Gowalla to filter out some inactive users in the chosen city.

\begin{table}[h]
	\centering
	\caption{Statistics of the experimental datasets.}\label{tab:statistics}	%显示表格的标题  
	\begin{tabular} {c|c|c|c} 
		\hline  
		Dataset & \#Users  & \#Locations & \#Check-ins \\		 %用&来分隔单元格的内容 
		\hline
		Gowalla & 1,270 & 16,615 & 67,194 \\
		\hline
		Foursquare & 1,092& 24,133 & 98,351 \\
		\hline
	\end{tabular} 
\end{table}

Through the above procedures of preprocessing, we have obtained our experimental datasets and their characteristics are shown in Table~\ref{tab:statistics}. 
For ease of testing, we randomly divide social relations into training sets, validation sets, and test sets with the ratios of 0.8, 0.1, and 0.1.
As is known to all, the number of negative social relation pairs is very large.
Therefore it is very time-consuming for evaluations.
For this consideration, we follow~\cite{Koren08,HeLZNHC17} by randomly selecting 50 users not linked with each target user as her/his candidates.
Finally, we have 18,304 and 19,128 relation pairs in the test sets of Gowalla and Foursquare, respectively. 

\subsection{Metrics}
The performance of user link inference in an LSN is measure by three widely used metrics, i.e. precision at position k (P@k), recall at position k (R@k), and area under the receiver operating characteristic curve (AUC).
Among them, P@k measures the faction of accurately linked users contained by returned top k candidates, and R@k denotes the faction of accurately estimated user links contained by all ground-truth links.
To be specific, P@k and R@k are given as follows:
\begin{equation}\label{eq:metric1}
    \PrecisionK = \frac{1}{Q_U}\sum_u\frac{N^u_{true}}{k}\,, ~~~   \RecallK = \frac{1}{Q_U}\sum_u\frac{N^u_{true}}{R_u}\,,
\end{equation}
where $N^u_{true}$ denotes the number accurately inferred links in top-K ranked candidates.
$R_u$ is the number of actual links in the ground truth set of user $u$.
$Q_U$ is the number of test users.
In the experiments, we set $k$ in $\PrecisionK$ and $\RecallK$ to 10 for comparison.

AUC relies on the true positive rate and false positive rate to compute its value.
It could be explained as the probability of ranking positive links higher than negative links.
The expectation value of AUC is equal to 0.5 for a random guess.
The computational details of AUC can be referred to~\cite{FengZWYZ0J19}.
For simplicity, we use Scikit-learn to obtain the metric scores.

\begin{figure*}[!t]
\centering
\begin{minipage}[c]{1\textwidth}
\subfigure[Performance comparison in AUC.]{\includegraphics[width=0.33\textwidth]{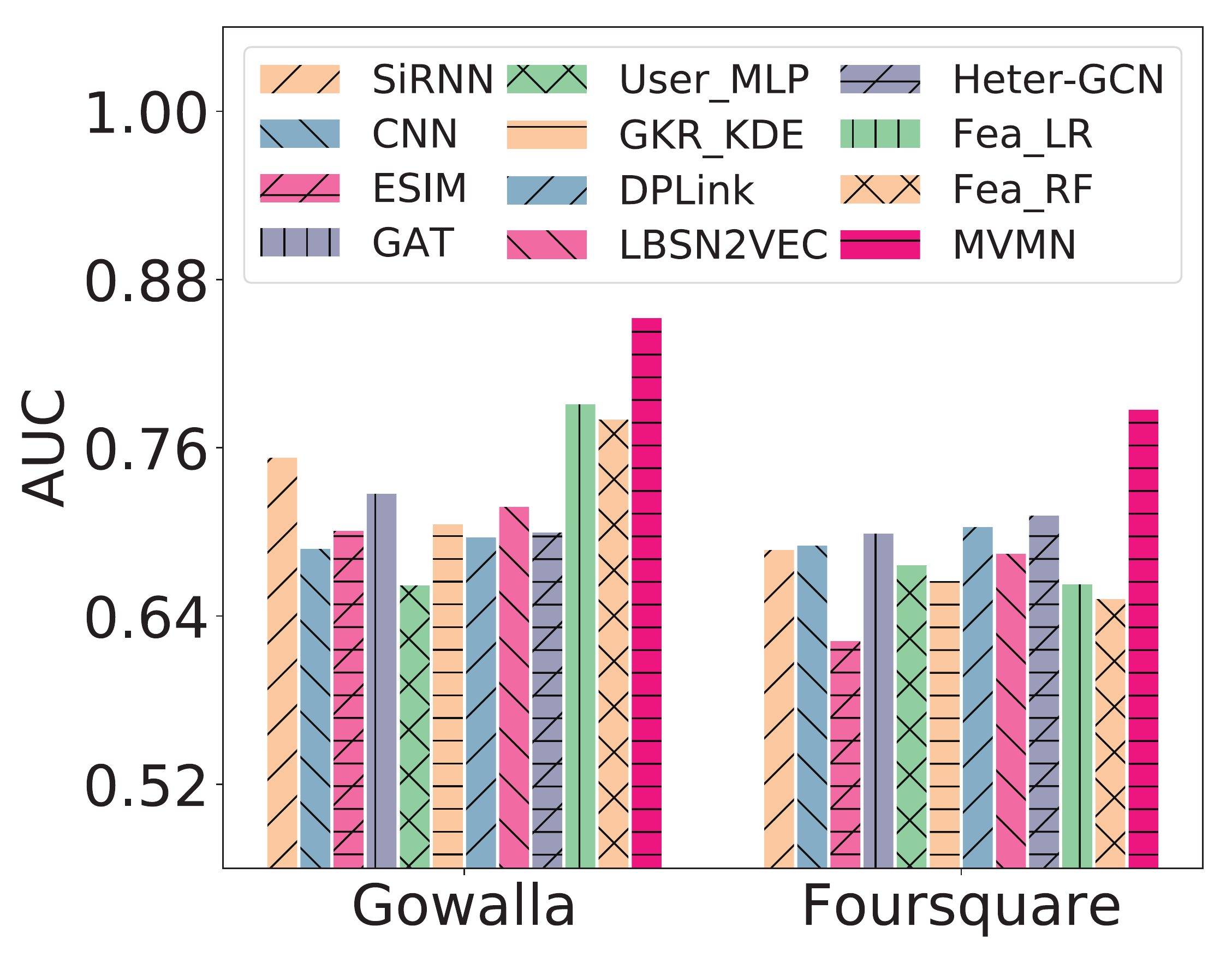}}
\subfigure[Performance comparison in R@10.]{\includegraphics[width=0.33\textwidth]{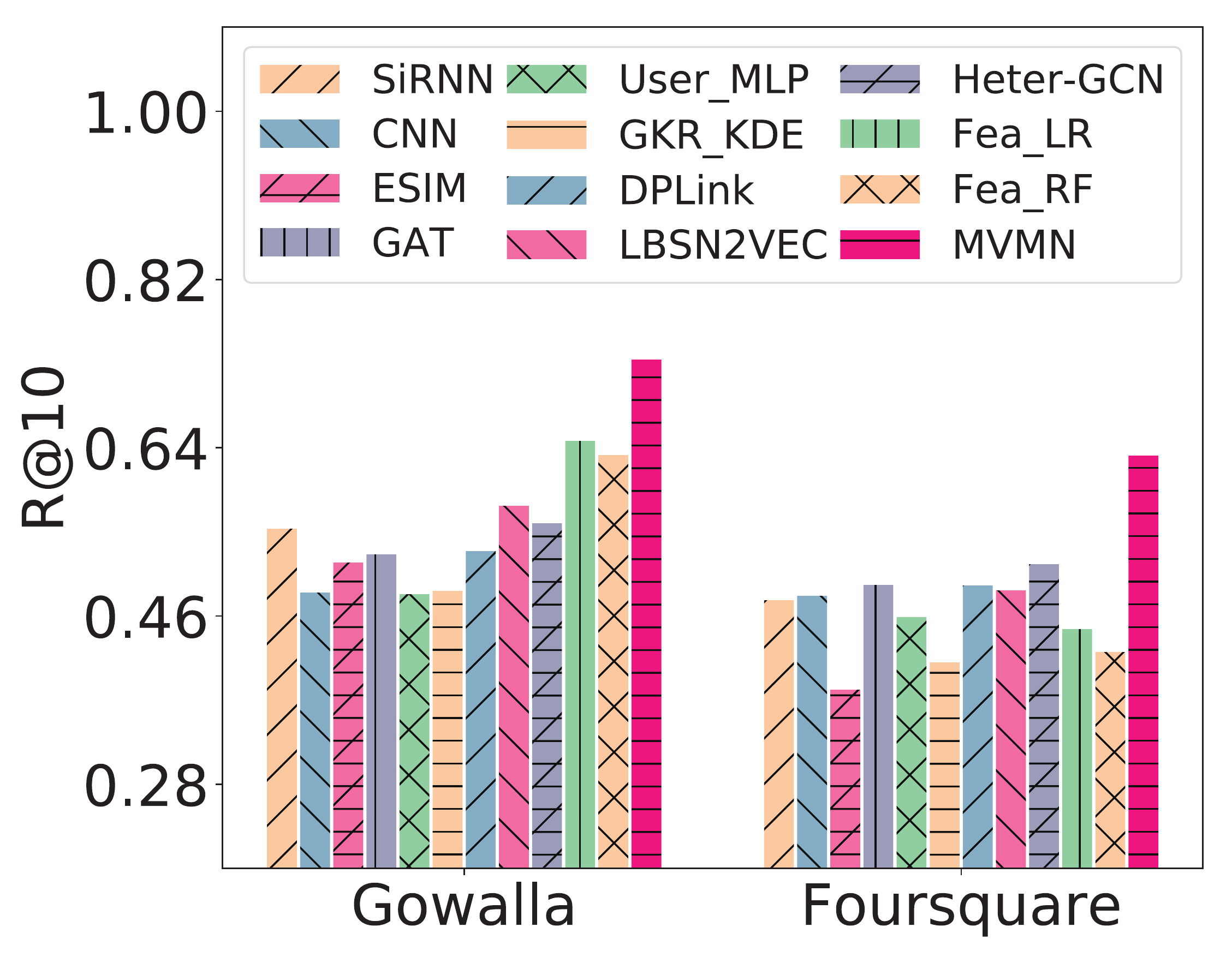}}
\subfigure[Performance comparison in P@10.]{\includegraphics[width=0.33\textwidth]{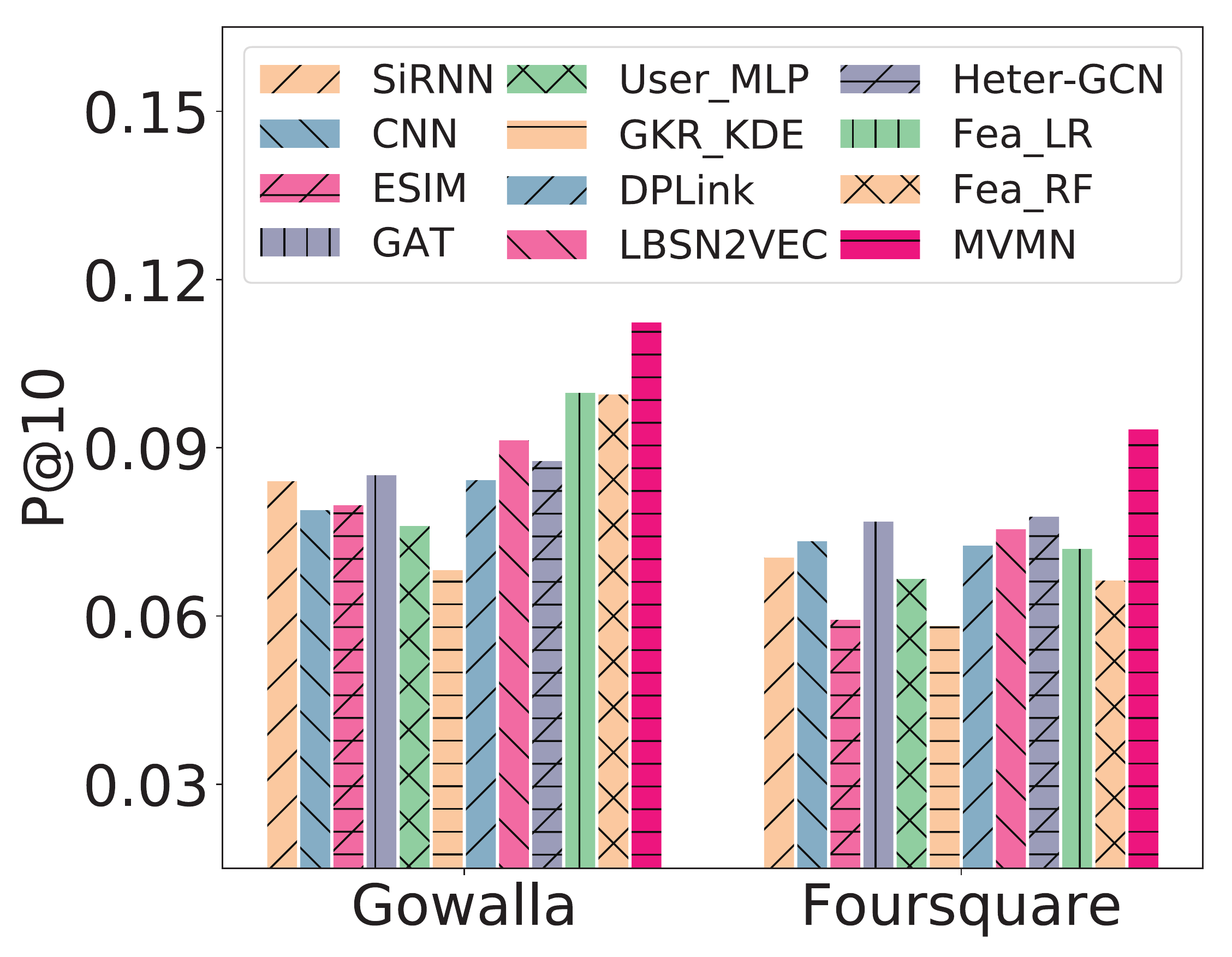}}

\caption{Evaluation results by our proposed model and baselines on the Gowalla and Foursquare datasets.}\label{fig:performance-comparison}
\end{minipage}
\end{figure*}

\subsection{Baselines}
We organize the baselines into four groups, according to the original problem each baseline focuses on.

\noindent\textbf{G1 --- Social link inference in an LSN:}

\textbf{Fea\_LR} and \textbf{Fea\_RF.} The pioneering study~\cite{Scellato-SIGKDD2011} adopts feature based classification methods.
Following this, we use logistic regression and random forest as the classification models, and name them as Fea\_LR and Fea\_RF, respectively.
The well-designed features, include social, temporal (the similarities mentioned in Section~\ref{sec:motivation}) and geographical features, are taken as the input of the two models.

\textbf{LBSN2Vec.} LBSN2Vec\footnote{\url{https://github.com/eXascaleInfolab/LBSN2Vec}}~\cite{YangQYC19} is a hypergraph embedding method which encodes user-user edges and user-time-POI-semantic hyperedges.
We adapt this method to our problem by only removing semantic nodes that might be missing in some domains and are not considered by this study. 

\textbf{Heter-GCN.} The model~\cite{WuLJC19} benefits from graph convolutional networks for learning social relations and recurrent neural networks to model trajectories.
The integration is achieved by initializing the graph node representations with the hidden representations obtained by RNNs.

\noindent\textbf{G2 --- User identity linkage across LSNs:}

\textbf{GKR\_GKE.} GKR\_GKE\footnote{\url{https://www.dropbox.com/s/mrkfao8gr8zktkw/ICDE-18-Chenwei.zip?dl=0}}~\cite{Chen-ICDE2018} is a kernel density based method which improves DG~\cite{Chen-CIKM2017} proposed earlier by the same authors. 

\textbf{DPLink.} 
Compared with other methods for user identity linkage, the novel insights of DBLink\footnote{\url{https://github.com/vonfeng/DPLink}}~\cite{FengZWYZ0J19} are the first appliance of deep learning techniques and the proposal of attention mechanism based selective matching.

\noindent\textbf{G3 --- Sequence matching:}

\textbf{SiRNN.} This is partially inspired by the study~\cite{MuellerT16} which proposes to use siamese RNNs for sentence matching.
The objective function for classification is based on cross-entropy loss, being consistent with ours.

\textbf{CNN.} Convolutional neural networks are utilized to measure the correlation of patient event sequences~\cite{Zhu-ICDM2016}.
We train this model in a supervised manner.

\textbf{ESIM.} ESIM\footnote{\url{https://github.com/lukecq1231/nli}}~\cite{ChenZLWJI17} builds on two recurrent neural networks and seizes the interaction between two sequence embeddings by its proposed interactive attention computation method.

\noindent\textbf{G4 --- General user recommendation:}

\textbf{User\_MLP.} We feed two user representations into multi-layer perceptrons (MLP) which correspond to the deep modeling part of NeuMF~\cite{HeLZNHC17}.

\textbf{GAT.} GAT~\cite{Velickovic-1710-10903} is directly applied to our problem by only considering social relations as edges and users as nodes.
A similar application can be found in the work~\cite{Wang0WFC19} which models a user-item bipartite graph.

\subsection{Implementations}
The hyperparameters of our model and baselines are tuned to achieve their better performance on the validation sets.
The maximal trajectory length $K$ is set to 200.
If a user's trajectory is longer than the threshold, we keep its recent 200 check-ins.
For the negative sampling adopted in model training mentioned in Section~\ref{subsec:multi-view}, we set the number of sampled users to 4.
By default, the embedding size is set to 64, the dimension of the hidden unit in RNNs (e.g., LSTM) is set to 128.
For the model GAT and our relation matching module, the number of attention header is equal to 3.
We use the Adam optimizer to train our model and other representation learning based methods.
The learning rate is 1e-4 and the batch size is 64.
We also consider using dropout with a ratio of 0.5 for intermediate layers to overcome the overfitting issue.
All the adopted models are run on a small server with two GPU GTX 1080Ti cards.
We implement our model by PyTorch 1.1.
If the baselines have source codes, we use them for comparison with some necessary minor modifications. Otherwise, we implement them.

\section{Experimental Results}\label{sec:er}
In this section, we conduct comprehensive comparisons to answer the following crucial research questions for a better understanding of our model, especially for the rationality of fusing multiple views for linking users:
\begin{itemize}%[leftmargin=1.8em]
	\item[\textbf{\texttt{Q1}:}] What is the performance of the proposed MVMN for link inference compared to strong competitors?
	
	\item[\textbf{\texttt{Q2}:}] Does each view-specific module in MVMN contribute to the final performance?
\end{itemize}

\subsection{Performance Comparison (Q1)}
Fig.~\ref{fig:performance-comparison} shows the performance of our model MVMN and other baselines, from which we observe the performance comparison is not exactly the same across the two datasets.

Among the baselines, we can observe that the feature based classification methods --- Fea\_LR and Fea\_RF achieve the best performance in the Gowalla dataset, and in the Foursquare dataset, Heter-GCN behaves a little better than other baselines.
This might be attributed to the reason that these methods within group G1 are tailored for the problem the same as ours.
LBSN2VEC is a little worse since its representation learning is not directly guided by optimizing social relations.
The other baselines have some issues when applied to our problem: (1) for the methods within group G4, only user relations are well utilized for inference; (2) the commonly adopted sequence matching models in G2 do not utilize temporal factors and social relations; and (3) the approaches belonging to group G3 do not address social relations and there is still room to improve the representation learning for trajectory matching.
Thus the observation meets our expectations to some extent. 

By further comparing the methods in group G1, we observe that the two feature based methods obtain relatively good results on Gowalla, significantly beating other baselines.
It might be the reason that the well-crafted features containing much domain-specific knowledge.
On the other hand, the graph representation learning based models obtain better results than the feature based methods on Foursquare.
This reflects the prospect of representation learning if it is well treated.

In the end, we can see the proposed MVMN outperforms all the competitors on both datasets.
In particular, on the Gowalla dataset, MVMN improves the state-of-the-art Fea\_LR method  from 0.647 to 0.734 in $\Recall10$, and from 0.100 to 0.112 in $\Precision10$.
Moreover, on the Foursquare dataset, MVMN improves the state-of-the-art Heter-GCN model from 0.515 to 0.631 in $\Recall10$, and from 0.078 to 0.093 in $\Precision10$.
We demonstrate later that the improvements are not only from the modeling of temporal dynamics based on temporal point process, but also thanks to the fusion of the learned matching representations from multiple views.

\subsection{Ablation Studies (Q2)}
\begin{table}[!t]
	\centering
	\caption{Ablation study of our model on the two datasets. }\label{tab:ablation} 	
	\begin{tabular} {c|cccc} 
		\toprule   
		Dataset & Method  & AUC & R@10 & P@10\\	
        \Xhline{0.7pt} 
		\multirow{5}*{Gowalla}   & MVMN & \textbf{0.852} & \textbf{0.734} & \textbf{0.112}  \\ \cline{2-5}
		& V1        & 0.826 & 0.709 & 0.105 \\ \cline{2-5}
		& V2        & 0.802 & 0.676 & 0.101 \\ \cline{2-5} 
        & V3        & 0.780 & 0.621 & 0.093 \\ \cline{2-5} 
		& V4        & 0.724 & 0.541 & 0.081 \\ 
		\Xhline{0.7pt} 
		\multirow{5}*{Foursquare}  & MVMN & \textbf{0.787} & \textbf{0.631} & \textbf{0.093} \\ \cline{2-5} 
		& V1     & 0.740 & 0.558 & 0.082 \\ \cline{2-5} 
		& V2     & 0.714 & 0.529 & 0.076 \\ \cline{2-5} 	
		& V3     & 0.681 & 0.492 & 0.079 \\ \cline{2-5} 
		& V4     & 0.699 & 0.503 & 0.075 \\ 
		\bottomrule  
	\end{tabular}  
\end{table}

We conduct ablation experiments to verify the contribution of each module.
This is a crucial step to understand how our model works.
To achieve this, we define the following variants of our model:
\begin{itemize}[leftmargin=*]
\item[*] \textbf{V1: MVMN w/o (RM)}. It removes the relation matching module from MVMN, which is equivalent to not explicitly modeling social relations in the model architecture. 

\item[*] \textbf{V2: MVMN w/o (RM+PP)}. To show how temporal point process influences our model, this method further erases $\Loss_p$ by setting $\beta$ to 0.
Noting that the temporal correlation computation of users in the time-series matching module is still maintained.

\item[*] \textbf{V3: MVMN w/o (RM+TM)}. This variant removes not only the relation matching module but also the time-series matching module.
It is equivalent to only using the location matching module for social link inference in a target LSN.

\item[*] \textbf{V4: MVMN w/o (RM+LM)}. It removes both the relation matching module and the location matching module.
Hence this variant only involves the time-series matching module.

\end{itemize}

Table~\ref{tab:ablation} presents the results of our full model and its variants, from which we have the following key observations:

%\vspace{0.2em} \noindent (1)  

\begin{enumerate}[leftmargin=*]
    \item 
The performance comparison between V1 and the full model MVMN demonstrates the positive contribution of the relation matching module.
The results are intuitive since social relations could reveal the user-to-user proximity from the aspect of network structures.
Moreover, compared to V1, both V3 and V4 incur performance drop.
This reflects the location matching module and the time-series matching module contribute to the final performance.
As such, the fusion of multiple views for matching is effective.

%\vspace{0.2em} \noindent (2)  
\item Compared to V1, V2 removes the point process loss $\Loss_p$, thereby only using twin LSTMs for modeling user temporal behavior sequences.
The results of V2 are significantly inferior to V1.
This indicates our model indeed benefits from the consideration of temporal point process.
\end{enumerate}

\begin{table}[!t]
	\centering
	\caption{Performance of alternative designs.}\label{tab:alternative} 		%显示表格的标题  
	\begin{tabular} {c|cccc} 
		\toprule   %第一行线  
		Dataset & Method  & AUC & R@10 & P@10 \\		 %用&来分隔单元格的内容 
        % \midrule   %第二行线
        \Xhline{0.7pt} 
		\multirow{7}*{Gowalla}  & V3      & 0.780 & 0.621 & 0.093 \\ %\cline{2-5} 
		                        & V3 w/ RNN                & 0.625 & 0.442 & 0.064 \\ \cline{2-5} 
		                        & V2      & 0.802 & 0.676 & 0.101 \\
	                        	& V3 w/ $\oplus$Time         & 0.778 & 0.619 & 0.094 \\ \cline{2-5}
	                        	& MVMN                  & 0.852 & 0.734 & 0.112 \\
	                            & MVMN w/ GCN                   & 0.818 & 0.694 & 0.107 \\
	                            & MVMN w/ Attention             & 0.822 & 0.705 & 0.106 \\ 
	   %	\midrule[0.6pt]
	   
		\Xhline{0.7pt} 
		\multirow{7}*{Foursquare}   & 
		V3    & 0.681 & 0.492 & 0.079 \\ %\cline{2-5} 
		                            & V3 w/ RNN               & 0.589 & 0.333 & 0.046 \\ \cline{2-5} 
		                            & V2      & 0.714 & 0.529 & 0.076 \\
	                            	&V3 w/ $\oplus$Time        & 0.688 & 0.509 & 0.079 \\ \cline{2-5} 
	                            	& MVMN                 & 0.787 & 0.631 & 0.093 \\
	                            	& MVMN w/ GCN                   & 0.732 & 0.553 & 0.081 \\ 
	                            	& MVMN w/ Attention            & 0.736 & 0.548 & 0.080 \\ 
		\bottomrule   %第三行线
	\end{tabular}  
\end{table}

\subsection{Alternative Designs for MVMN}\label{subsec:alternatives}

We design some alternatives of MVMN and its variants to better understand the intuition behind the proposed model:
\begin{itemize}[leftmargin=*]
\item[*] \textbf{V3 w/ RNN} uses RNNs in V3 to learn location representations for location-to-location correlation calculation.

\item[*] \textbf{V3 w/ $\oplus$Time} concatenates temporal embeddings with location embeddings in V3 for correlation calculation, in contrast to using LSTMs for representing temporal sequences in V2.  

\item[*] \textbf{MVMN w/ GCN} leverages GCN instead of GAT for learning social relations.

\item[*] \textbf{MVMN w/ Attention} performs attention computation to determine the importance weights of each view and further fuse them with weighted combination.

\end{itemize}

Table~\ref{tab:alternative} shows the results of the alternatives, from which we find:

%\vspace{0.2em} \noindent (1)  
\begin{enumerate}[leftmargin=*]
\item V3 w/ RNN degrades the performance of link inference across the two datasets when using RNN modeling.

%\vspace{0.2em} \noindent (2)  
\item Compared to V2, V3 w/ $\oplus$Time performs worse in most cases.
It indicates that using LSTMs for modeling temporal sequences is more effective than tightly fusing  spatial and temporal information.
As such, it might be better to attribute spatial and temporal information to different views.

%\vspace{0.2em} \noindent (3)  
\item MVMN w/ GCN is inferior to MVMN using GAT, showing the benefit of automatically quantifying the weights of social relations.
Besides, adopting attention computation to fuse the representations of different views does not improve the performance.

\end{enumerate}

\subsection{More Discussions}

\subsubsection{Effect of $\beta$}
\begin{figure}[!t]
\centering
\begin{minipage}[c]{0.38\textwidth}
\includegraphics[width=1\textwidth]{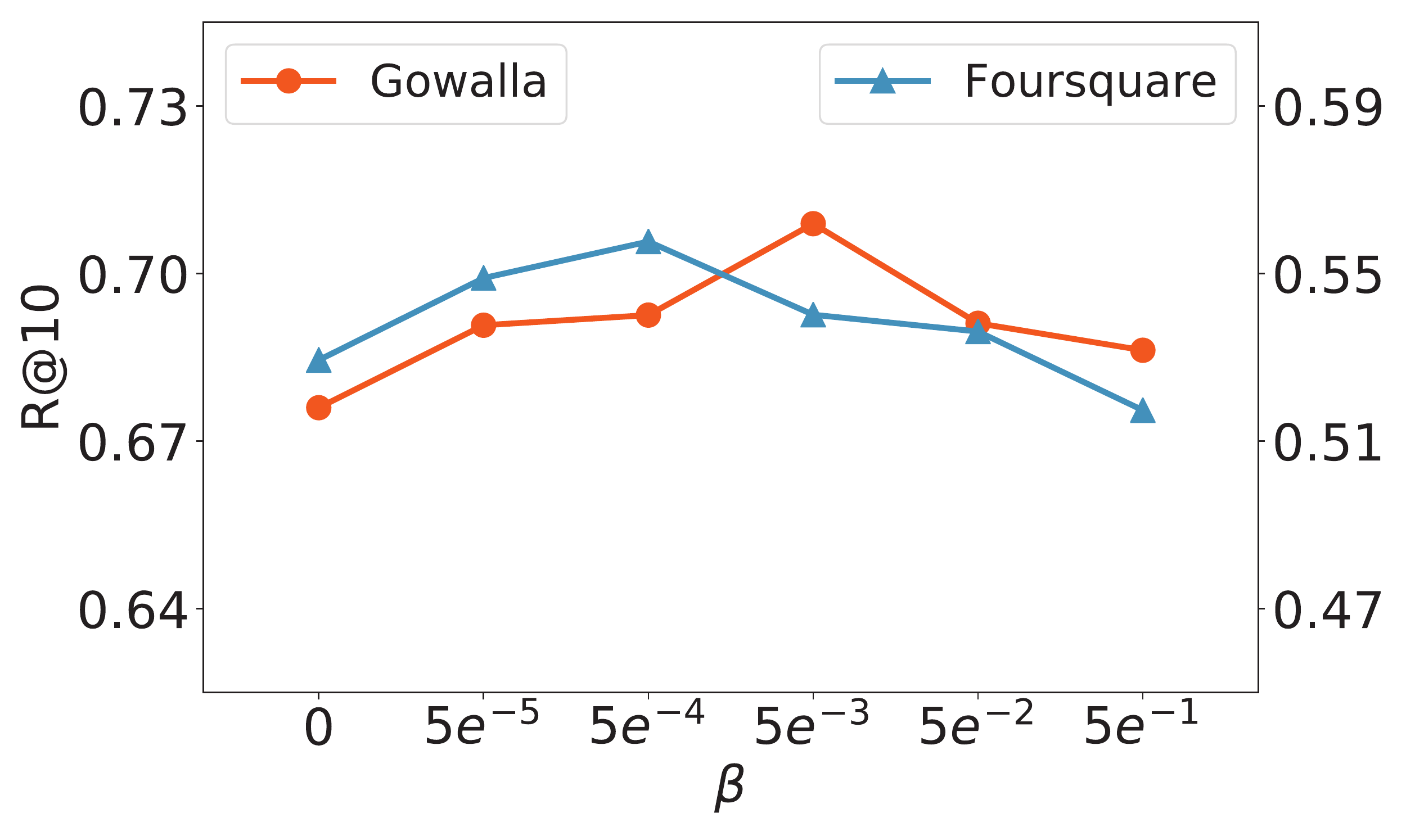}
\caption{Performance variation with different $\beta$.}\label{fig:beta}
\end{minipage}
\end{figure}

We show how the performance of link inference changes with different values of $\beta$ in Fig.~\ref{fig:beta}.
The results corresponding to $\beta=0$ are also reported for ease of illustration.
Through comparison, we observe that when $\beta$ takes values larger than 0, the results are much better than the results with $\beta=0$.
This phenomenon indicates the consideration of temporal point process is beneficial for the studied problem.

\subsubsection{Effect of propagation layer number}

\begin{figure}[!t]
\centering
\begin{minipage}[c]{0.38\textwidth}
\includegraphics[width=1\textwidth]{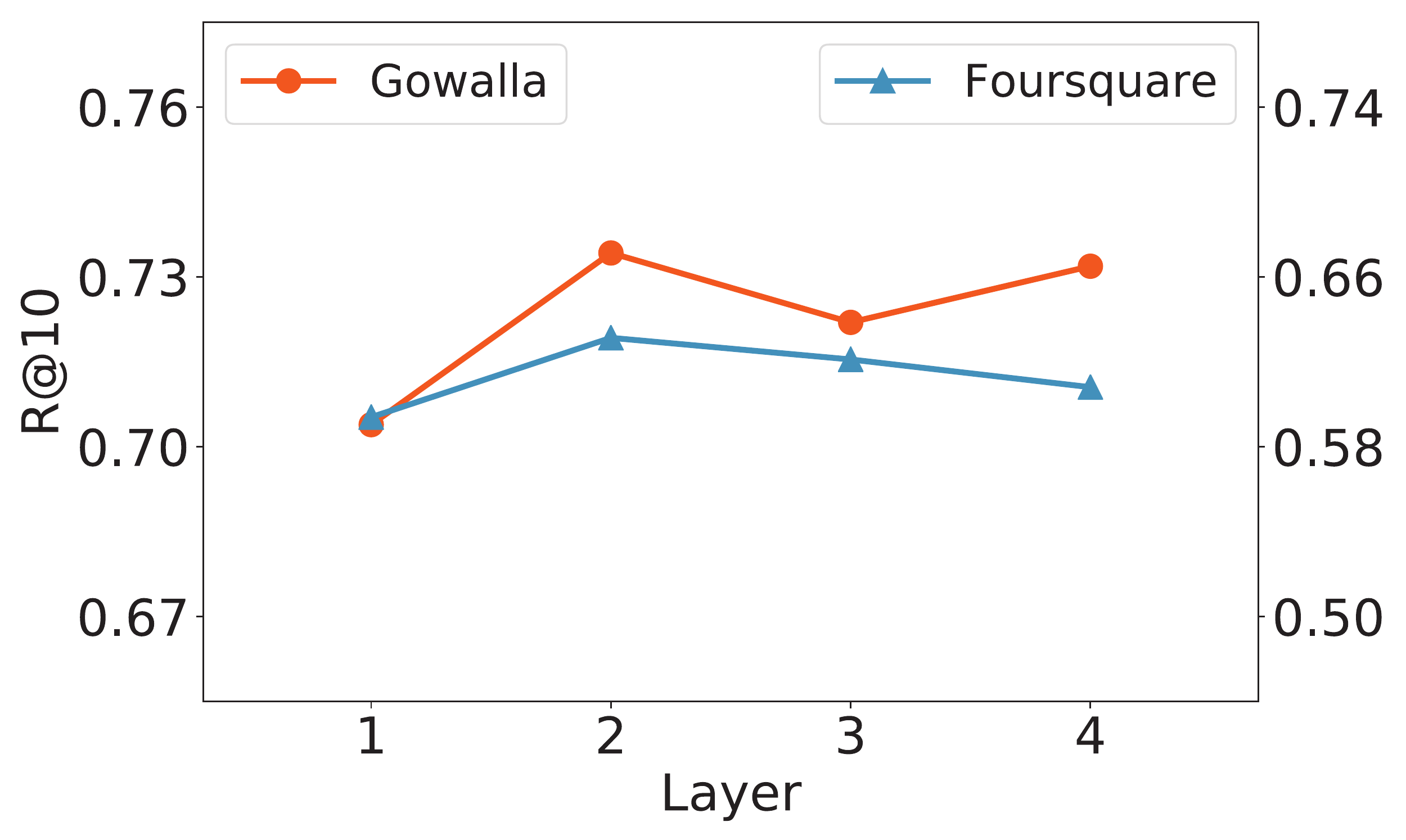}
\caption{Performance variation with different layer number in representation propagation in GNN.}\label{fig:layer}
\end{minipage}
\end{figure}

Fig.~\ref{fig:layer} depicts how the user linking performance is affected by the number of representation propagation layers. 
We can see the results are improved with a larger number of propagation layers at the beginning, and consequently, the results remain to be relatively stable.
This phenomenon reveals that learning high-order user relations beyond existing direct relation brings additional advantages.

\section{Conclusion}\label{sec:con}
In this paper, we have investigated the problem of social link inference in a given location-aware social network by treating spatial, temporal, and social factors of user pairs as different views for effective user linking.
We have devised the multi-view matching network which is welcomed for the benefit of learning the view-specific representation by each matching module.
Recurrent neural temporal point process is incorporated into our model to guide the effective representation learning of user time-series.
We have conducted extensive experiments and demonstrated our model consistently outperforms strong competitors proposed for this problem and related ones.
The rationality of our model design has also been validated by ablation studies. 

%%%%%%%%%%%%%%%%%%%%%%%%%%%%%%%%%%%%%%%%%%%%%%%%%%%%%%%%%%%%%%%%%%%%%%%%%%%%%%%%%%%%%%

% or


%\appendices

%\section{}

%\section*{Acknowledgment}


\ifCLASSOPTIONcaptionsoff
  \newpage
\fi



%\bibliographystyle{IEEEtran}
\bibliographystyle{IEEEtran}
\bibliography{reference}


\begin{IEEEbiography}[{\includegraphics[width=1.1in,height=1.35in,clip,keepaspectratio]{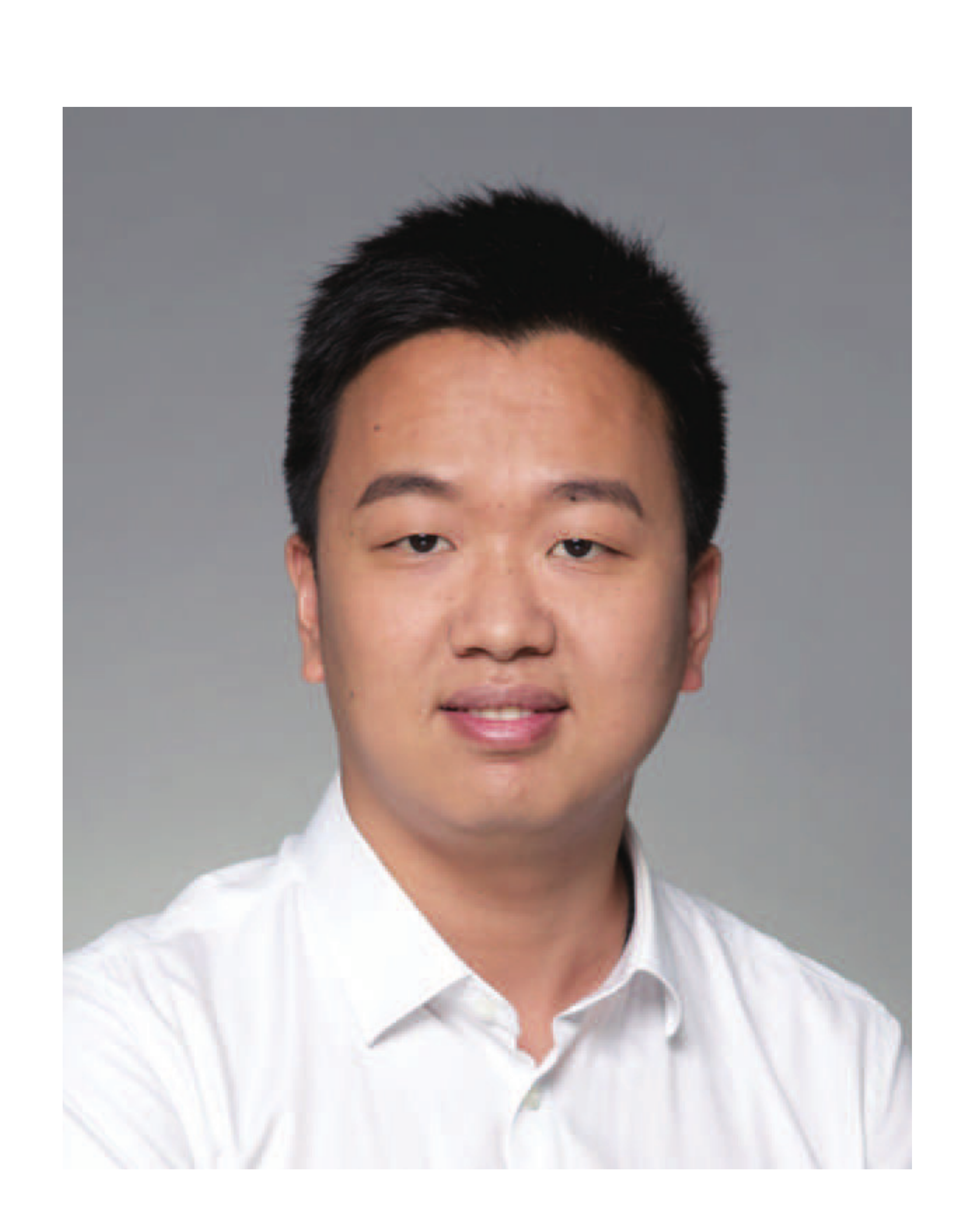}}]{Wei Zhang} received his PHD degree in computer science and technology from Tsinghua university, Beijing, China, in 2016. He is currently an associate researcher in the School of Computer Science and Technology, East China Normal University, Shanghai, China.
His research interests mainly include data mining and machine learning applications, especially for user generated data modeling.
He has published more than 30 papers at some leading international conferences and journals.
He served as an area chair for ICDM, a PC member for some leading international conferences, such as SIGKDD, SIGIR, WWW, and an invited reviewer for top journals like IEEE TKDE, IEEE TNNLS, ACM TKDD, etc.
\end{IEEEbiography}

\begin{IEEEbiography}[{\includegraphics[width=1.1in,height=1.35in,clip,keepaspectratio]{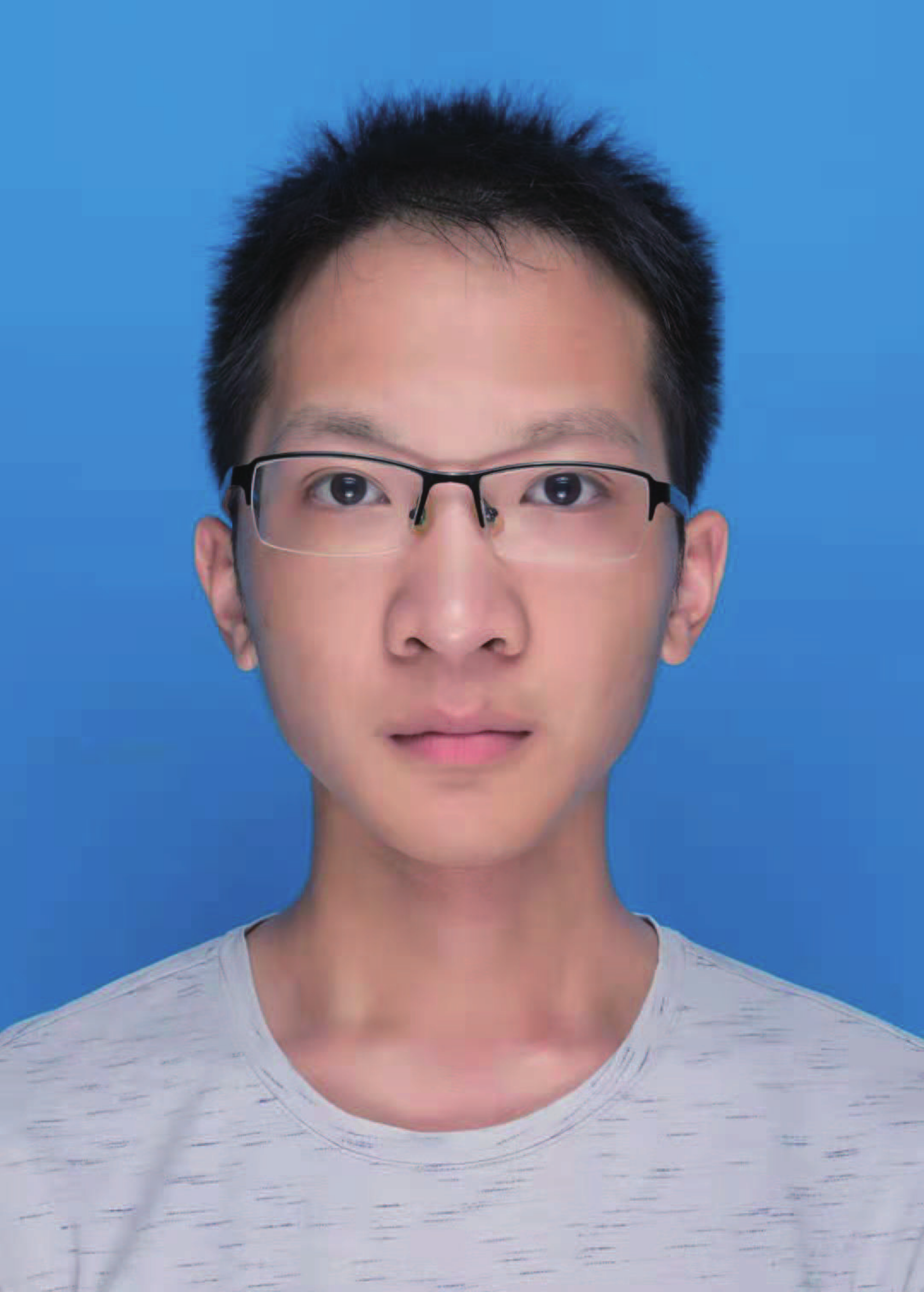}}]{Xin Lai} received the B.Sc. degree in computer science and technology from Xiangtan University, Xiangtan, China, in 2018. He is currently pursuing the master degree with the Department of Computer Science and Technology, East China Normal University, Shanghai. His main research area is sequential user behavior modeling.
\end{IEEEbiography}

%\newpage

\begin{IEEEbiography}[{\includegraphics[width=1.1in,height=1.35in,clip,keepaspectratio]{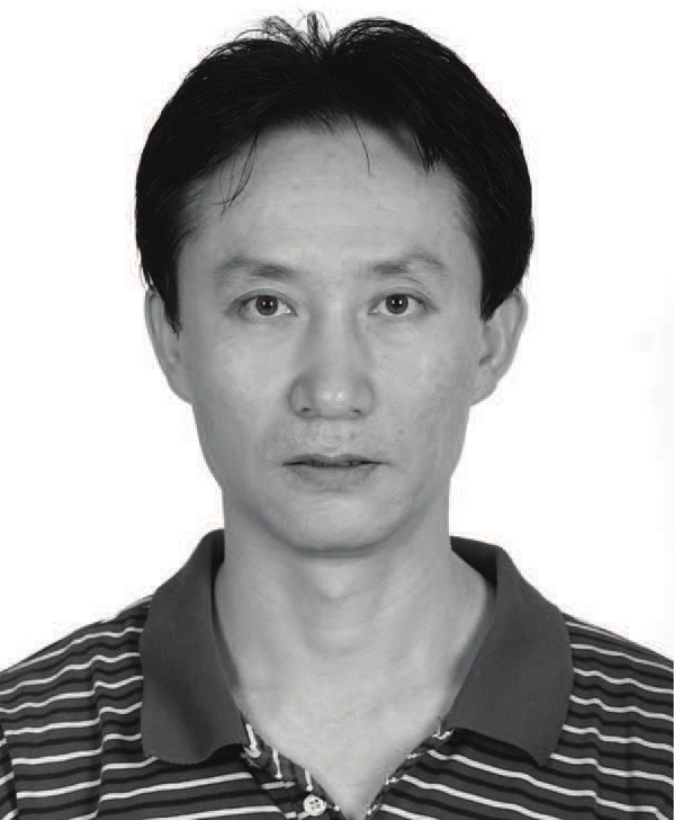}}]{Jianyong Wang} received the PhD degree in computer science from the Institute of Computing Technology, Chinese Academy of Sciences, in 1999. He is currently a professor in the Department of Computer Science and Technology, Tsinghua University, Beijing, China. He was an assistant professor with Peking University, and visited Simon Fraser University, the University of Illinois at Urbana-Champaign, and the University of Minnesota at Twin Cities before joining Tsinghua University in December 2004. His research interests mainly include data mining and Web information management. He has co-authored more than 60 papers in some leading international conferences and some top international journals. He is serving or has served as a PC member for some leading international conferences, such as SIGKDD, VLDB, ICDE, WWW, and an associate editor of the IEEE Transactions on Knowledge and Data Engineering and the ACM Transactions on Knowledge Discovery from Data. He is a fellow of the IEEE.
\end{IEEEbiography}


%\vfill


\end{document}